# Recommending Clinical Trials for Online Patient Cases using Artificial Intelligence

**Joey Chan, BS[1], Qiao Jin, MD[1], Nicholas Wan, BE[1], Charalampos S. Floudas, MD[2], Elisabetta Xue, MD[2], Zhiyong Lu, PhD[1]**

[1]National Library of Medicine, National Institutes of Health, Bethesda, MD;
[2]National Cancer Institute, National Institutes of Health, Bethesda, MD

**Abstract**

*Clinical trials are crucial for assessing new treatments; however, recruitment challenges—such as limited awareness, complex eligibility criteria, and referral barriers—hinder their success. With the growth of online platforms, patients, caregivers, and family members increasingly post medical cases on social media and health communities, while physicians publish case reports accessible on platforms like PubMed—collectively expanding recruitment pools beyond traditional clinical trial pathways. Recognizing this potential, we utilized TrialGPT, a framework that leverages a large language model, to match 50 online patient cases (collected from case reports and social media) to clinical trials and evaluate performance against traditional keyword-based searches. Our results show that TrialGPT outperformed traditional methods by 46%, with patients eligible, on average, for 7 of the top 10 recommended trials. Additionally, outreach to case authors and trial organizers yielded positive feedback. These findings highlight TrialGPT's potential to expand patient access to specialized care through non-traditional sources.*

## 1 Introduction

Recruitment of patients to clinical trials remains a significant challenge that directly impacts trial success. Around 80% of clinical trials terminate due to the failure to meet enrollment targets within the expected timeframe, leading to wasted resources, delayed medical advancements, and limited treatment options for patients in need[1]. The lack of participation from patients in clinical trials can be attributed to multiple factors, including limited awareness and understanding of clinical research, restrictive trial eligibility criteria, geographic and financial barriers, mistrust in medical institutions, concerns about potential risks and side effects, and low physician recommendations. However, the most significant barrier remains lack of awareness, as in 2020, only 9.4% of U.S. adults had full awareness of clinical trials, while 41% knew nothing about them[2,3].

Discovering suitable clinical trials is often complex, and patients rely on healthcare providers as their most trusted source of information. Getz (2017) found that 84% of patients would join a trial if referred by their physician[4-6]. However, physicians face barriers such as limited time, insufficient trial information, and resource constraints, leading to missed enrollment opportunities. Consequently, many patients—especially those at facilities without active trials—must search for options independently. This can be overwhelming due to the complexity of medical information and eligibility criteria.

Increasingly, patients turn to online platforms like Reddit health communities and Facebook, where they connect with others facing similar conditions[31]. Social media is particularly valuable for those with rare diseases, cancer, or chronic conditions, where standard treatments are often ineffective. Miller et al. found that many with rare diseases seek social media support due to limited treatment options[19]. Currently, 300 million people worldwide live with rare diseases, yet 95% lack treatments approved by the U.S. Food and Drug Administration[21]. Similarly, 80% of cancer patients use social media for support and treatment exploration, with 49% citing diagnosis as their reason for joining, particularly in aggressive cases like glioblastoma[18,22]. Those with chronic conditions also engage in peer discussions to navigate self-management and emerging treatments[20]. Beyond patient discussions, healthcare providers often share case reports on rare conditions and treatment outcomes[28,35]. However, as patient cases are shared across multiple platforms in diverse formats, traditional clinical trial search methods struggle to keep pace, making it increasingly difficult to identify relevant trials for those outside standard treatment protocols.

We recently proposed TrialGPT[9], which leverages a large language model (LLM) as its backbone to match patients with suitable clinical trials. TrialGPT, which has been evaluated with 183 synthetic patients across more than 75,000 trial eligibility annotations, demonstrated a reduced screening time by 42.6% in patient recruitment and high accuracy in patient-to-trial matching, achieving 87.3% accuracy in criterion-level eligibility predictions, closely aligning with expert performance.

To further validate its effectiveness with real-world data, we conducted an evaluation using 50 online patient cases sourced from published case reports in PubMed and three online health communities on Reddit. These cases were matched to clinical trials using TrialGPT and compared against results using the traditional keyword search method. In total, TrialGPT matched 500 patient-trial pairs, while the traditional keyword search method produced 475 pairs—yielding a combined total of 975 patient-trial pairs. The traditional keyword search method had 25 fewer matches because two online patient cases returned no clinical trials, and one case returned only 5. Then, two graduate-level researchers with biomedical knowledge manually evaluated the patient-trial pairs, and two internal clinicians selected the most promising clinical trials from these evaluations for outreach to the trial organizer and case authors—specifically for the PubMed case reports. This process details our study pipeline visualized in Figure 1. Our findings indicate that TrialGPT significantly outperformed traditional approaches, achieving an average eligibility rate of 75.7% compared to 29.7%. These results highlight TrialGPT's potential to enhance clinical trial matching, particularly for patients with complex medical conditions.

## 2 Methods

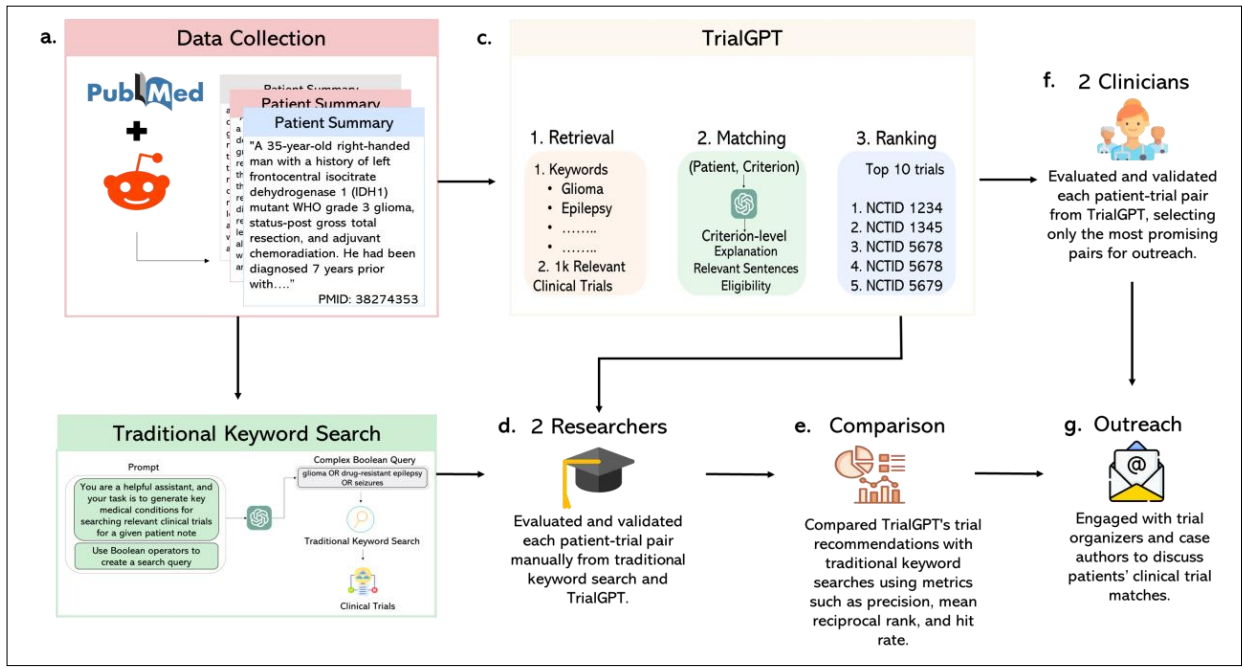

**Figure 1.** Study design. This figure illustrates the workflow for the curation and evaluation of clinical trial matches. **a.** Online patient cases published in 2024 on PubMed and posted on Reddit communities (AskDocs, Rare Diseases, and Cancer) were collected. **b.** The traditional keyword search method used the collected cases as input to prompt GPT-4o, which generated a complex Boolean query based on patient conditions. This query was then used in the traditional clinical trial search, resulting in a list of clinical trials for each patient. **c.** The collected cases were also used as input for TrialGPT, which consists of three components — retrieval, matching, and ranking — resulting in a ranked list clinical trials for each patient. **d.** Two graduate-level researchers with biomedical knowledge reviewed and validated the patient–trial pairs recommended by both TrialGPT and the traditional keyword search method. **e.** Both methods were compared using Precision, Mean Reciprocal Rank to assess eligibility, and Hit Rate to evaluate the identification of beneficial trials. **f.** Two internal clinicians, using their domain expertise, selected the most promising patient–trial pairs recommended by TrialGPT from the PubMed case reports for outreach. **g.** Case authors and trial organizers of the selected patient–trial pairs were contacted to assess match eligibility, trial benefit, and the clarity of TrialGPT explanations.

### 2.1 Datasets

*Patient Summaries Data*. To compose the 50 online patient cases, we collected patient summaries from various sources to ensure a diverse and representative dataset. We gathered 25 of the most recent patient summaries from case reports published on PubMed, selecting cases from January 2024 to August 2024 that described patients in the United States with ongoing symptoms after treatment. Additionally, we sourced 25 Reddit posts: 15 from the AskDocs community,

5 from a Cancer community, and 5 from a Rare Diseases community, all posted between May 2024 and August 2024. Table 1 provides an overview of the cases from different sources, highlighting key characteristics such as sex distribution, age distribution, and average case length. Notably, the patient summaries in our dataset are significantly longer than the synthetic summaries from TrialGPT's original dataset—averaging approximately 469 words compared to 118 words, which were often brief. Moreover, while posts from Reddit communities tended to be verbose, the PubMed case reports provided more extensive and relevant medical information.

**Table 1.** Statistics of online patient cases used in this study. Variables are reported as mean ± standard deviation where applicable. We report the number of cases (N), sex distribution, patient age, and case length (in words) across PubMed case reports and Reddit communities.

| Data source | Case reports | Reddit – AskDocs | Reddit – Rare Diseases | Reddit – Cancer |
|---|---|---|---|---|
| **N** | 25 | 15 | 5 | 5 |
| **Sex (male: female)** | 14: 11 | 4: 11 | 1: 4 | 2: 3 |
| **Age (year)** | 39 ± 23.6 | 33.4 ± 18.2 | 33.4 ± 19.0 | 34.4 ± 14.4 |
| **Case length (word)** | 602.0 ± 272.5 | 430.1 ± 142.5 | 319.8 ± 212.2 | 527 ± 208.3 |

*Clinical Trials Data.* We gathered data on approximately 23,000 actively recruiting clinical trials in the United States, including both interventional and observational studies, sourced from ClinicalTrials.gov (CT.gov) in July 2024. These trials were filtered solely by location, considering only those that are actively recruiting in the United States. Unlike many other countries that rely on multiple national or regional registries, the United States primarily uses CT.gov as its centralized database for registering and reporting clinical trials. This makes it the most comprehensive and consistent source for clinical trial data within a single country, ensuring data completeness and comparability. In contrast, trials conducted in other countries are often registered across multiple platforms, such as the WHO International Clinical Trials Registry Platform (ICTRP), EU Clinical Trials Register (EU-CTR), or national registries, making it difficult to aggregate and analyze data systematically[30]. For each trial, we collected its Brief Title, Phase, Drugs, Drugs List, Diseases, Diseases List, Enrollment, Inclusion Criteria, Exclusion Criteria, and Brief Summary.

## 2.2 TrialGPT

In this study, we leveraged GPT-4o (version 2024-02-01), accessed through Microsoft Azure's OpenAI services, with the inference temperature fixed at 0 to ensure deterministic outputs. GPT-4o served as the backbone LLM for TrialGPT, which consists of three key components: TrialGPT-Retrieval, TrialGPT-Matching, and TrialGPT-Ranking. First, TrialGPT-Retrieval identifies relevant clinical trials using a combination of keyword generation and hybrid retrieval methods. The LLM extracts up to 32 relevant keywords from both the patient note and a summarized version of the note, which are then processed by a hybrid retrieval system that integrates both BM25[37] for lexical search and MedCPT[10] for semantic search to retrieve the most relevant clinical trials. The top 1,000 trials from this process advance to the next stage. Next, TrialGPT-Matching conducts a detailed criterion-by-criterion analysis, comparing the patient summary against the inclusion and exclusion criteria of the retrieved clinical trials. It identifies relevant sentences from the patient summary for each criterion and provides explanations for the match. Finally, TrialGPT-Ranking orders the matched trials based on their scores. This ranking aggregates scores from TrialGPT-Matching's criterion-by-criterion analysis, along with relevance and eligibility scores assigned by the LLM, ensuring that the most relevant trials are prioritized. In this study, the top 10 highest-ranked clinical trials are selected, resulting in a total of 500 patient-to-trial pairs. We refer interested readers to Jin et al. (2024) for more details about TrialGPT[9].

## 2.3 Evaluation

*Evaluation Metrics.* Our study evaluates patient-to-trial eligibility as the primary outcome and whether a trial is beneficial for patients as the secondary outcome, where a trial is evaluated for its benefit only if the patient is first deemed eligible for that trial. A patient is considered eligible if they meet all inclusion criteria and none of the exclusion criteria. A trial is deemed beneficial only if it provides a controlled treatment assessment and access to novel therapies that offer direct patient benefits. This includes the systematic evaluation of treatment efficacy, targeted therapies, or advanced medical strategies that are not yet available in standard clinical practice, thereby improving disease management and patient prognosis[29]. We report eligibility using Precision for recommended trials at ranks 1, 3, 5, and 10, as well as the Mean Reciprocal Rank (MRR). Precision at rank $k$ is defined as the proportion of eligible trials within the top $k$ recommendations and can be computed as:

$$\text{P@k} = \frac{\text{Number of eligible trials in top k}}{k} \quad (1)$$

MRR measures how early the first eligible clinical trial appears in the ranked list and is defined as the average of the reciprocal ranks of the first relevant result for each patient case, which is computed as:

$$\text{MRR} = \frac{1}{N}\sum_{i=1}^{N}\frac{1}{\text{rank}_i} \quad (2)$$

where $N$ is the total number of patient cases evaluated, and $\text{rank}_i$ is the position of the first eligible clinical trial in the ranked list for the $i^{\text{th}}$ patient case.

We measured the Hit Rate at each trial rank, denoted by $[t_1, t_2,\ldots,t_{10}]$, by calculating, for each trial position, the proportion of patient cases for which a beneficial clinical trial was found at or before that specific trial rank. The Hit Rate at trial rank $t$ is computed as:

$$\text{HitRate@t} = \frac{1}{N}\sum_{i=1}^{N}\mathbf{1}(\text{a beneficial trial found by rank } t \text{ for patient } i) \quad (3)$$

where $N$ is the total number of patient cases evaluated, and $\mathbf{1}(\cdot)$ is an indicator function equal to 1 if a beneficial trial was found by trial rank $t$ for patient $i$, and 0 otherwise.

These metrics comprehensively assess TrialGPT's effectiveness in identifying relevant clinical trials, offering a quantitative measure of how efficiently and accurately the model can assist patients and clinicians in finding suitable trial opportunities.

*Manual Evaluation.* Two graduate-level researchers with biomedical knowledge evaluated the trials recommended by TrialGPT for the dataset of 50 online patient cases, resulting in 500 patient-trial matchings. One researcher assessed each patient summary against the eligibility criteria for each matched trial, determining whether the patient met all inclusion criteria and none of the exclusion criteria. If the trial was deemed eligible, the researcher then evaluated its potential benefit for the patient based on their medical condition. The second researcher validated these results to ensure agreement. The inter-rater agreement was 96.4%, and any disagreements were resolved through discussion.

## 2.4 Outreach

Two internal clinicians were recruited to further evaluate and re-rank the eligible cases assessed by the two researchers on the PubMed case reports. They selected only the most promising and beneficial patient-trial pairs for outreach consideration. The clinicians were provided with patient summaries along with the clinical trials recommended by the researchers. They assessed whether the patient was eligible for the trial and whether the trial would provide meaningful medical benefits. Then, they determined whether outreach was necessary, which was defined as cases where the patient was both eligible and likely to benefit from enrollment, taking into account of additional considerations such as patient location and accessibility.

For each patient-trial pair, both the case author — who is part of the healthcare team caring for the patient and presenting the case — and the trial organizers representing the clinical trials were contacted based on the clinicians' outreach responses[23]. Specifically, only the case authors of the PubMed patient summaries were contacted if clinicians deemed outreach necessary, while only the trial organizers of the corresponding trials were contacted when warranted. Both groups were engaged to gain a comprehensive perspective on TrialGPT's performance. Contact information for case authors was obtained from the PubMed author information section, while trial organizers were identified through CT.gov. Their responses were recorded in a standardized form, with a follow-up sent one week after the initial contact. If no response was received within two weeks, the case was marked as unresponsive.

The outreach aimed to assess three key aspects: (1) patient-to-trial eligibility, (2) whether the clinical trial would helpful/benefit the patient (measured on a Likert scale from 1 to 5), and (3) the clarity of TrialGPT's explanation for each criterion (also measured on a Likert scale from 1 to 5). The same set of questions was asked to both case authors and trial organizers to enable a direct comparison of their perspectives. Case authors were provided with the clinical trials that our internal clinicians had deemed eligible and beneficial, along with TrialGPT's criterion-by-criterion analysis explaining the reasoning behind each match. Similarly, trial organizers received the online patient case along with TrialGPT's reasoning, identical to what was provided to case authors but specific to their trial.

### 2.5 Comparing TrialGPT with Traditional Keyword Search Method

*Traditional Keyword Search Method.* We define traditional keyword search method as using CT.gov to identify relevant clinical trials. CT.gov is a database comprising over 400,000 registered studies[14]. Traditional keyword search consists of two approaches: basic and advanced. Since TrialGPT utilizes an LLM, we ensured a fair comparison by leveraging GPT-4o with zero-shot prompting to generate complex Boolean queries for the CT.gov conditions field, restricting results to actively recruiting trials in the United States as we had done for the clinical trials used in TrialGPT. This approach involved extracting relevant keywords from patient summaries, specifically focusing on relevant medical conditions.

*Evaluation.* Like our evaluation technique for TrialGPT, the same two researchers manually assessed the top 10 clinical trials retrieved through the traditional keyword search method for each online patient case, resulting in 475 patient-to-trial matches. The researchers applied the same evaluation criteria—eligibility and benefit—to the patient-trial pairs. The inter-rater agreement was 94.3%. To evaluate eligibility performance, we used Precision and Mean Reciprocal Rank to assess how accurately and effectively the system ranked eligible trials. For beneficial trials, we calculated the Hit Rate across the top 10 clinical trials recommended. This approach ensured consistency with the evaluation framework used for TrialGPT and provided a comprehensive assessment of clinical trial relevance.

## 3 Results

### 3.1 Overall Eligibility Performance

In this section, we compare the clinical trial recommendations by TrialGPT with those using the traditional keyword search method across the dataset, categorized by data source. A comparative analysis is presented in Table 2. We assessed patient eligibility for trials using Precision at ranks 1, 3, 5, and 10, where rank *n* refers to the top *n* recommended trials for a given patient. Additionally, we evaluated performance using the MRR. Upon examination, TrialGPT consistently outperformed the traditional keyword search method by a substantial margin across all metrics.

**Table 2.** Eligibility Performance of TrialGPT and the Traditional Keyword Search Method. P@k refers to Precision at rank *k*, measuring the proportion of eligible trials within the top *k* recommendations. MRR indicates how highly the first eligible trial appears in the ranking, with higher values reflecting earlier correct matches. Values are reported per data source.

| Method/Metric | Data source | P@1 | P@3 | P@5 | P@10 | MRR |
|---|---|---|---|---|---|---|
| Traditional Keyword Search | Case reports | 0.440 | 0.266 | 0.232 | 0.208 | 0.520 |
| | Reddit – AskDocs | 0.333 | 0.312 | 0.320 | 0.280 | 0.457 |
| | Reddit – Rare Diseases | 0.400 | 0.467 | 0.360 | 0.500 | 0.600 |
| | Reddit – Cancer | 0.200 | 0.268 | 0.200 | 0.200 | 0.300 |
| | **Overall Average** | **0.343** | **0.328** | **0.278** | **0.297** | **0.469** |
| TrialGPT | Case reports | 0.880 | 0.850 | 0.840 | 0.820 | 0.930 |
| | Reddit – AskDocs | 0.725 | 0.746 | 0.736 | 0.748 | 0.800 |
| | Reddit – Rare Diseases | 0.800 | 0.934 | 0.920 | 0.760 | 0.900 |
| | Reddit – Cancer | 1.000 | 0.868 | 0.800 | 0.700 | 1.000 |
| | **Overall Average** | **0.851** | **0.849** | **0.824** | **0.757** | **0.975** |

More specifically, the average eligibility performance of TrialGPT surpasses that of the traditional keyword search method across all evaluated metrics. TrialGPT achieved an average improvement of 50.8% in Precision at rank 1 (P@1), 52.1% at rank 3 (P@3), 54.6% at rank 5 (P@5), and 46.0% at rank 10 (P@10). This means that patients are, on average, eligible for approximately 7 out of the top 10 recommended trials with TrialGPT, compared to only about 3 out of 10 with the traditional keyword search method. These results demonstrate a significant improvement over the traditional keyword search method, indicating that the process of identifying eligible clinical trials for patients is greatly enhanced with the use of TrialGPT. Additionally, TrialGPT showed a 50.1% improvement in MRR. While the traditional keyword search method typically presents the first eligible trial around the third result, TrialGPT, on average, ranks the first eligible trial at the very top of the list. This advantage is particularly crucial for patients with complex medical conditions, where timely access to appropriate clinical trials can significantly impact treatment decisions and outcomes. TrialGPT not only identifies more eligible trials but also ranks them higher, ensuring that patients and healthcare providers can quickly find the most relevant options without needing to sift through less pertinent results — a crucial advantage when every moment counts in guiding care and improving patient outcomes.

### 3.2 Evaluating Beneficial Outcomes in Top 10 Clinical Trials

In this section, we evaluate the performance of TrialGPT across the top 10 recommended clinical trials in comparison to the traditional keyword search method. Specifically, this analysis focuses on our secondary outcome: the proportion of beneficial trials identified among the top 10 trials for each patient. Performance was assessed using the Hit Rate, averaged across patients and stratified by data source. The results of this analysis are presented in Figure 2.

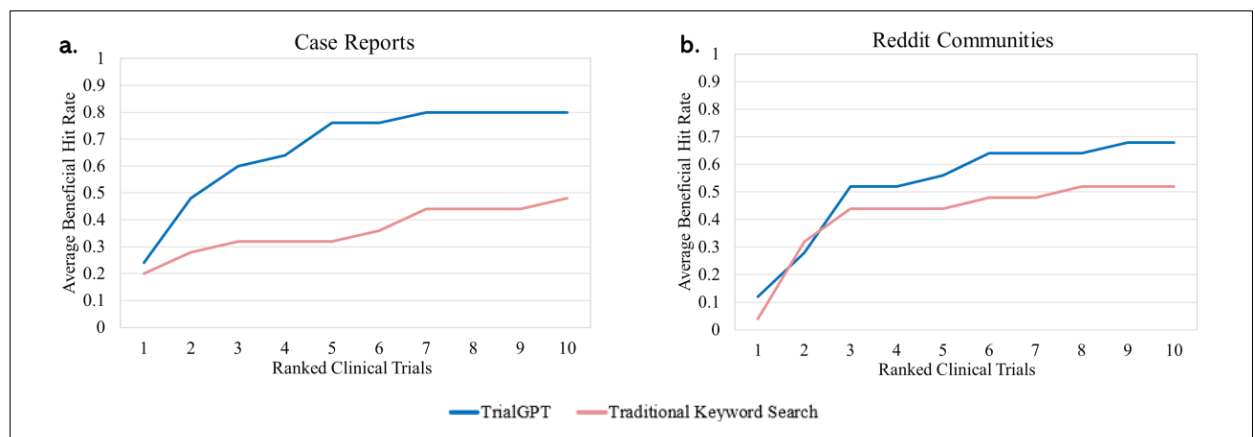

**Figure 2.** Comparison of Hit Rate performance for identifying beneficial clinical trials using TrialGPT and the traditional keyword search method. **a.** Average beneficial Hit Rate across the top 10 recommended clinical trials for PubMed case reports. **b.** Average beneficial Hit Rate across the top 10 recommended clinical trials for Reddit communities (AskDocs, Rare Diseases, and Cancer).

Our findings indicate that TrialGPT is more effective than the traditional keyword search method in identifying beneficial clinical trials for patients across both data sources, as shown in Figure 2. Specifically, Figure 2a illustrates that TrialGPT outperformed the traditional keyword search method in identifying beneficial clinical trials from PubMed case reports, supporting our hypothesis that the presence of more precise medical terminology allows TrialGPT to perform better. In contrast, Figure 2b shows that the traditional keyword search method performed slightly better in identifying beneficial clinical trials from online patient cases from Reddit communities.

Overall, TrialGPT outperformed the traditional search method in identifying beneficial clinical trials across the entire dataset. When evaluating trial-by-trial performance for the PubMed case reports, we found that by the 10th recommended trial, TrialGPT achieved an average Hit Rate of 80%, compared to only 48% for the traditional keyword search method. In other words, patients using TrialGPT encountered a beneficial clinical trial within the first 10 recommendations 80% of the time, whereas those relying on the traditional keyword search method did so only 48% of the time. Similarly, in the Reddit posts, TrialGPT achieved a 68% Hit Rate, while the traditional search method reached only 52% by the 10th recommended trial. Although the difference in Reddit-based cases is smaller than in the PubMed case reports, this 16% improvement is substantial. It highlights that TrialGPT significantly increases the likelihood that patients — even those describing their conditions in informal, less structured language on platforms like Reddit — can quickly identify clinical trials that may have a direct impact on their care. This is especially important as an increasing number of patients turn to social media platforms and online health communities for medical guidance. Moreover, it is important to note that the recommended clinical trials from the traditional search method in this study relied on prompting an LLM to generate complex Boolean queries from patient summaries — a level of refinement that may not be practical or commonly used by patients or healthcare providers in real-world manual search scenarios. These findings demonstrate that TrialGPT is robust across both structured, provider-generated patient narratives and informal, patient-generated content, making it highly applicable to a wide range of real-world use cases.

### 3.3 Outreach Analysis

This section examines outreach responses from case authors and trial organizers contacted based on patient-trial pairs that clinicians identified as the most promising from TrialGPT's clinical trial recommendations for the PubMed case reports. Outreach was conducted for 20 patient-trial pairs, with 20 trial organizers and 12 case authors contacted. More trial organizers were reached out to than case authors because some patients had multiple suitable clinical trials. Their perspectives on TrialGPT's patient-to-trial matching recommendations offer insight into its role in clinical decision-making and trial selection interpretability. Although the response rate was low—8% for case authors and 15% for trial organizers—this can be attributed to several limitations, which we detail in Section 5. Nevertheless, these responses highlight the system's practical impact from two key perspectives: healthcare providers and trial organizers. Below, we present the perspectives of the case authors and trial organizers detailing how they evaluated the relevance and clarity of TrialGPT's recommendations.

*Case Authors' Perspective.* Case authors who evaluated TrialGPT's recommendations found their recommended trial or trials helpful and confirmed patient eligibility, noting that the explanations provided for each criterion were clear and informative. They also recognized TrialGPT's ability to effectively explain its rationale, further reinforcing its usefulness in patient-to-trial matching.

*Trial Organizers' Perspective.* Trial organizers also reviewed TrialGPT's recommendations for different cases, providing insights into its role in refining patient enrollment decisions. While some found the recommended trials highly relevant and the eligibility explanations exceptionally clear, others, despite confirming patient eligibility, indicated that the trial's overall helpfulness to the patient was low. In cases where a patient was not eligible for their trial, they still found the criterion explanations structured and informative.

Across both perspectives, TrialGPT received positive feedback for its ability to enhance interpretability and support decision-making in the trial selection process. Its detailed criterion-based explanations not only improved transparency but also helped both case authors and trial organizers better understand the rationale behind trial recommendations.

## 4 Related Works

### 4.1 Traditional Keyword Search for Clinical Trial Matching

The challenges of clinical trial matching have been widely studied in the past, with various approaches developed to improve the identification of eligible trials for patients. Traditional clinical trial search methods, including basic and advanced keyword search, often present challenges for both patients and healthcare providers. Basic keyword search returns numerous irrelevant results, requiring significant time to filter through, while advanced search relies on Boolean operators, demanding technical expertise. Even experienced users can spend over 100 hours refining search strategies[11,15]. While these methods provide access to comprehensive trial databases, studies have shown that they are often ineffective at identifying suitable trials for patients without specialized knowledge.

### 4.2 Artificial Intelligence-Driven Clinical Trial Matching Systems

To improve efficiency and reduce manual review, studies have adopted Artificial Intelligence (AI) for clinical trial matching. Early rule-based systems used regex to filter ineligible patients but struggled with unstructured clinical notes due to variability and lack of standardization[24]. To address this, ML and NLP techniques, including Named Entity Recognition (NER) and ontology-based normalization (SNOMED-CT, UMLS), were introduced. Hybrid rule-based and ML models improved accuracy but struggled with contextual understanding, negations, and temporal constraints[25]. The introduction of LLMs has significantly improved semantic comprehension in trial matching[36]. Key contributors include frameworks such as TrialGPT, PRISM, and Cognitive Technology for Cancer Trial Matching[26,27]; systems like criteria2query[32]; and models including DeepEnroll and COMPOSE[33,34]. In this work, we demonstrate that TrialGPT outperforms the traditional keyword search method by achieving higher eligibility precision and identifying more beneficial trials, particularly for a dataset that is composed of real-world diverse online patient cases that haven't ever been explored by clinical matching systems. Outreach results further highlight strong positive feedback from trial organizers and case authors, underscoring the promise of LLM-driven patient recruitment.

## 5 Discussion and Conclusions

In summary, this study successfully utilized TrialGPT to match online patient cases with relevant clinical trials. To the best of our knowledge, it is the first study to leverage such diverse online patient cases for trial matching, creating new opportunities for clinical trial recruitment that benefit both patients and healthcare providers. Our findings demonstrate that TrialGPT significantly outperforms the traditional keyword search method in identifying eligible trials, with improvements of 46%, and on average, patients are eligible for 7 of the top 10 recommended trials. This improvement is particularly evident for patients with rare diseases, cancer, and chronic conditions, where existing trial search mechanisms often fail to provide relevant matches.

Despite these promising results, several limitations must be acknowledged. First, the manual collection of online patient cases proved to be a time-intensive process. Additionally, case reports often exhibit publication bias, as case authors are more likely to publish rare diseases and successful treatment outcomes rather than unsuccessful treatments or persistent symptoms. This bias presents a challenge in identifying patient cases that accurately reflect the broader spectrum of real-world patient experiences.

Furthermore, although online patient cases often contained more detailed narratives compared to the synthetic patient cases used in TrialGPT evaluation, they frequently lacked critical medical details. This limitation made it difficult to fully assess whether a patient met all clinical trial eligibility criteria, particularly for trials with complex inclusion and exclusion requirements. Given the ethical considerations of this study, the inability to contact patients directly restricted opportunities to supplement missing information, further limiting the depth of clinical assessment. Future work could explore collaborations with healthcare institutions to facilitate follow-up with patients, allowing for a more comprehensive collection of clinical details while ensuring ethical compliance.

Another key limitation lies in the scope of online patient cases included in this study. While we examined cases from PubMed and Reddit, we were unable to incorporate a wider range of patient communities, such as PatientsLikeMe, Facebook, or X. Expanding to these platforms could have provided a broader representation of patient narratives and offered insights into how different online communities engage with clinical trial information. Various platforms attract distinct patient populations based on factors such as disease type, condition severity, and access to healthcare resources. Understanding these variations is crucial for determining how TrialGPT can be applied to diverse patient populations. Future research should further investigate these engagement patterns to refine TrialGPT's ability to match patients across a broader range of online health communities.

Additionally, our study focused on outreach to case authors and trial organizers to evaluate TrialGPT's ability to determine patient eligibility and identify beneficial clinical trials. Consequently, only the 25 online patient cases from PubMed case reports were evaluated in detail by clinicians—who themselves often face heavy workloads. A low response rate from both groups made it challenging to comprehensively assess TrialGPT's performance in a real-world clinical context. Case authors, often occupied with demanding clinical and research responsibilities, may not be the primary physician for the patient but rather a member of the care team. As a result, they might be less involved or may have already moved on, and the patient's status might have changed. Additionally, it is uncommon for a corresponding author to receive such a request; they may have considered it spam or been less responsive, particularly since TrialGPT was applied to patient narratives written after its publication, making trial recommendations less of a priority. In contrast, trial organizers showed a slightly higher response rate, likely driven by their ongoing need for patient recruitment. These discrepancies highlight the importance of patients having timely access to clinical trial opportunities and the challenges they face when navigating the process independently. Many patients, especially those with rare diseases or complex conditions, struggle to identify suitable trials due to the complexity of eligibility criteria and limited guidance from healthcare providers who may not always be aware of available options or have the time to discuss these options with them. As online platforms become an increasingly common source of medical information and support, this study has demonstrated that even when using patient summaries from online platforms that may not always provide complete clinical details, tools like TrialGPT can still help bridge this gap by improving accessibility and enabling patients to connect with relevant clinical trials more effectively.

In conclusion, this study highlights the feasibility and potential impact of using AI-driven models like TrialGPT to streamline clinical trial matching for a broader scope of patient cases—extending beyond traditional clinical notes or Electronic Health Records (EHRs). Specifically, our focus on both clinician-derived patient narratives and patient-driven data demonstrates how these alternative sources of information can be leveraged to identify relevant trials. By automating patient-to-trial identification and providing transparent, criterion-level explanations for eligibility, TrialGPT serves as a valuable tool in bridging the gap between patients and clinical research. Moving forward, further validation and enhancements to TrialGPT will be crucial for optimizing its role in clinical trial recruitment, ultimately improving patient access to treatments and advancing medical research.

**Acknowledgements**   We thank the case authors and trial organizers for their valuable feedback. This research was supported by the Division of Intramural Research (DIR) of the National Library of Medicine (NLM), National Institutes of Health (NIH), and 2024 NIH Director's Challenge Innovation Award.

## References

1. Brøgger-Mikkelsen M, Ali Z, Zibert JR, Andersen AD, Thomsen SF. Online Patient Recruitment in Clinical Trials: Systematic Review and Meta-Analysis. Journal of Medical Internet Research [Internet]. 2020 Sep 15;22(11):e22179. Available from: https://doi.org/10.2196/22179
2. Leiter A, Diefenbach MA, Doucette J, Oh WK, Galsky MD. Clinical trial awareness: Changes over time and sociodemographic disparities. Clinical Trials [Internet]. 2015 Feb 10;12(3):215–23. Available from: https://doi.org/10.1177/1740774515571917
3. Yadav S, Todd A, Patel K, Tabriz AA, Nguyen O, Turner K, et al. Public knowledge and information sources for clinical trials among adults in the USA: evidence from a Health Information National Trends Survey in 2020. Clinical Medicine [Internet]. 2022 Sep 1;22(5):416–22. Available from: https://doi.org/10.7861/clinmed.2022-0107
4. Getz KA. US physician and nurse proclivity to refer their patients into clinical trials. Therapeutic Innovation & Regulatory Science [Internet]. 2020 Jan 6;54(2):404–10. Available from: https://doi.org/10.1007/s43441-019-00069-3
5. Getz KA. Examining and enabling the role of health care providers as patient engagement facilitators in clinical trials. Clinical Therapeutics [Internet]. 2017 Oct 25;39(11):2203–13. Available from: https://doi.org/10.1016/j.clinthera.2017.09.014
6. Getz KA. Examining and enabling the role of health care providers as patient engagement facilitators in clinical trials. Clinical Therapeutics [Internet]. 2017 Oct 25;39(11):2203–13. Available from: https://doi.org/10.1016/j.clinthera.2017.09.014
7. Ridgeway JL, Asiedu GB, Carroll K, Tenney M, Jatoi A, Breitkopf CR. Patient and family member perspectives on searching for cancer clinical trials: A qualitative interview study. Patient Education and Counseling [Internet]. 2016 Aug 24;100(2):349–54. Available from: https://doi.org/10.1016/j.pec.2016.08.020
8. Schindler TM, Grieger F, Zak A, Rorig R, Konka KC, Ellsworth A, et al. Patient preferences when searching for clinical trials and adherence of study records to ClinicalTrials.gov guidance in key registry data fields. PLoS ONE [Internet]. 2020 May 29;15(5):e0233294. Available from: https://doi.org/10.1371/journal.pone.0233294
9. Jin Q, Wang Z, Floudas CS, Chen F, Gong C, Bracken-Clarke D, et al. Matching patients to clinical trials with large language models. Nature Communications [Internet]. 2024 Nov 18;15(1). Available from: https://doi.org/10.1038/s41467-024-53081-z
10. Jin Q, Kim W, Chen Q, Comeau DC, Yeganova L, Wilbur WJ, et al. MedCPT: Contrastive Pre-trained Transformers with large-scale PubMed search logs for zero-shot biomedical information retrieval. Bioinformatics [Internet]. 2023 Nov 1;39(11). Available from: https://doi.org/10.1093/bioinformatics/btad651
11. Glanville JM, Duffy S, McCool R, Varley D. Searching ClinicalTrials.gov and the International Clinical Trials Registry Platform to inform systematic reviews: what are the optimal search approaches? Journal of the Medical Library Association JMLA [Internet]. 2014 Jul 1;102(3):177–83. Available from: https://doi.org/10.3163/1536-5050.102.3.007
12. Li YH, Li YL, Wei MY, Li GY. Innovation and challenges of artificial intelligence technology in personalized healthcare. Scientific Reports [Internet]. 2024 Aug 16;14(1). Available from: https://doi.org/10.1038/s41598-024-70073-7
13. Liu H, Chi Y, Butler A, Sun Y, Weng C. A knowledge base of clinical trial eligibility criteria. Journal of Biomedical Informatics [Internet]. 2021 Apr 1;117:103771. Available from: https://doi.org/10.1016/j.jbi.2021.103771
14. Gresham G, Meinert JL, Gresham AG, Piantadosi S, Meinert CL. Update on the clinical trial landscape: analysis of ClinicalTrials.gov registration data, 2000–2020. Trials [Internet]. 2022 Oct 6;23(1). Available from: https://doi.org/10.1186/s13063-022-06569-2
15. Bramer WM, De Jonge GB, Rethlefsen ML, Mast F, Kleijnen J. A systematic approach to searching: an efficient and complete method to develop literature searches. Journal of the Medical Library Association JMLA [Internet]. 2018 Oct 4;106(4). Available from: https://doi.org/10.5195/jmla.2018.283
16. Kalankesh LR, Monaghesh E. Utilization of EHRs for clinical trials: a systematic review. BMC Medical Research Methodology [Internet]. 2024 Mar 18;24(1). Available from: https://doi.org/10.1186/s12874-024-02177-7
17. Thompson MA. Social media in clinical trials. American Society of Clinical Oncology Educational Book [Internet]. 2014 May 1;(34):e101–5. Available from: https://doi.org/10.14694/edbook_am.2014.34.e101
18. Fridman I, Bylund CL, Lafata JE. Trust of social media content and risk of making misinformed decisions: Survey of people affected by cancer and their caregivers. PEC Innovation [Internet]. 2024 Aug 17;5:100332. Available from: https://doi.org/10.1016/j.pecinn.2024.100332

19. Miller EG, Woodward AL, Flinchum G, Young JL, Tabor HK, Halley MC. Opportunities and pitfalls of social media research in rare genetic diseases: a systematic review. Genetics in Medicine [Internet]. 2021 Jul 19;23(12):2250–9. Available from: https://doi.org/10.1038/s41436-021-01273-z
20. Alhusseini N, Banta JE, Oh J, Montgomery SB. Social media use for health purposes by chronic disease patients in the United States. Saudi Journal of Medicine and Medical Sciences [Internet]. 2020 Dec 26;9(1):51–8. Available from: https://doi.org/10.4103/sjmms.sjmms_262_20
21. Health NLG. The landscape for rare diseases in 2024. The Lancet Global Health [Internet]. 2024 Feb 15;12(3):e341. Available from: https://doi.org/10.1016/s2214-109x(24)00056-1
22. Lamborn KR, Chang SM, Prados MD. Prognostic factors for survival of patients with glioblastoma: Recursive partitioning analysis. Neuro-Oncology [Internet]. 2004 Jul 1;6(3):227–35. Available from: https://doi.org/10.1215/s1152851703000620
23. Agrawal A, Eiger D, Jain D, Allman R, Eiger G. The right to write: Who "Owns" the case report? European Journal of Case Reports in Internal Medicine [Internet]. 2019 Jan 16;(Vol 6 No 1):1. Available from: https://pmc.ncbi.nlm.nih.gov/articles/PMC6372049/
24. Bickell NA, May B, Havrylchuk I, John J, Lin S, Tao A, et al. Implementation of a rule-based algorithm to find patients eligible for cancer clinical trials. JAMIA Open [Internet]. 2024 Oct 8;7(4). Available from: https://doi.org/10.1093/jamiaopen/ooae131
25. Sharif MT, Rehman A. Systematic Literature Review on Clinical trial eligibility Matching [Internet]. arXiv.org. 2025. Available from: https://arxiv.org/abs/2503.00863
26. Beck JT, Vinegra M, Dankwa-Mullan I, Torres A, Simmons CC, Holtzen H, et al. Cognitive technology addressing optimal cancer clinical trial matching and protocol feasibility in a community cancer practice. Journal of Clinical Oncology [Internet]. 2017 May 20;35(15_suppl):6501. Available from: https://doi.org/10.1200/jco.2017.35.15_suppl.6501
27. Gupta SK, Basu A, Nievas M, Thomas J, Wolfrath N, Ramamurthi A, et al. PRISM: Patient Records Interpretation for Semantic Clinical Trial Matching using Large Language Models. arXiv (Cornell University) [Internet]. 2024 Apr 23; Available from: https://arxiv.org/abs/2404.15549
28. Suvvari TK. Are case reports valuable? Exploring their role in evidence based medicine and patient care. World Journal of Clinical Cases [Internet]. 2024 Jul 10;12(24):5452–5. Available from: https://doi.org/10.12998/wjcc.v12.i24.5452
29. Gilmartin-Thomas JF, Liew D, Hopper I. Observational studies and their utility for practice. Australian Prescriber [Internet]. 2018 May 31;41(3):82–5. Available from: https://doi.org/10.18773/austprescr.2018.017
30. Hoffmann JM, Grossmann R, Widmann A. Academic clinical trials: Publication of study results on an international registry—We can do better! Frontiers in Medicine [Internet]. 2022 Nov 24;9. Available from: https://doi.org/10.3389/fmed.2022.1069933
31. Chirumamilla S, Gulati M. Patient Education and Engagement through Social Media. Current Cardiology Reviews [Internet]. 2019 Nov 22;17(2):137–43. Available from: https://doi.org/10.2174/1573403x15666191120115107
32. Yuan C, Ryan PB, Ta C, Guo Y, Li Z, Hardin J, et al. Criteria2Query: a natural language interface to clinical databases for cohort definition. Journal of the American Medical Informatics Association [Internet]. 2018 Nov 29;26(4):294–305. Available from: https://doi.org/10.1093/jamia/ocy178
33. Zhang X, Xiao C, Glass LM, Sun J. DeepEnroll: Patient-Trial Matching with Deep Embedding and Entailment Prediction. arXiv (Cornell University) [Internet]. 2020 Jan 1; Available from: https://arxiv.org/abs/2001.08179
34. Gao J, Xiao C, Glass LM, Sun J. COMPOSE: Cross-Modal Pseudo-Siamese Network for Patient Trial Matching. arXiv (Cornell University) [Internet]. 2020 Jan 1; Available from: https://arxiv.org/abs/2006.08765
35. Zhao Z, Jin Q, Chen F, Peng T, Yu S. A large-scale dataset of patient summaries for retrieval-based clinical decision support systems. Scientific Data [Internet]. 2023 Dec 18;10(1). Available from: https://doi.org/10.1038/s41597-023-02814-8
36. Tian S, Jin Q, Yeganova L, Lai PT, Zhu Q, Chen X, et al. Opportunities and challenges for ChatGPT and large language models in biomedicine and health. Briefings in Bioinformatics [Internet]. 2023 Nov 22;25(1). Available from: https://doi.org/10.1093/bib/bbad493
37. Robertson S, Zaragoza H. The Probabilistic Relevance Framework: BM25 and beyond. Foundations and Trends® in Information Retrieval [Internet]. 2009 Jan 1;3(4):333–89. Available from: https://doi.org/10.1561/1500000019